%% file: main_arxiv.tex
\documentclass[aps,prl,twocolumn,superscriptaddress,nobalancelastpage]{revtex4-2}
\usepackage{iftex}
\usepackage[russian,english]{babel}
\usepackage{preamble}

\begin{document}
\title{Quasiprobability Thermodynamic Uncertainty Relation} 
\author{Kohei Yoshimura}
\email{kohei.yoshimura@riken.jp}
\affiliation{Nonequilibrium Quantum Statistical Mechanics RIKEN Hakubi Research Team, Pioneering Research Institute (PRI), RIKEN, 2-1 Hirosawa, Wako, Saitama 351-0198, Japan}
\affiliation{Universal Biology Institute, The University of Tokyo, 7-3-1 Hongo, Bunkyo-ku, Tokyo 113-0033, Japan}
\author{Ryusuke Hamazaki}
\affiliation{Nonequilibrium Quantum Statistical Mechanics RIKEN Hakubi Research Team, Pioneering Research Institute (PRI), RIKEN, 2-1 Hirosawa, Wako, Saitama 351-0198, Japan}
\affiliation{RIKEN Center for Interdisciplinary Theoretical and Mathematical Sciences (iTHEMS), RIKEN, 2-1 Hirosawa, Wako, Saitama 351-0198, Japan}
\date{\today}

\begin{abstract}
    We derive a quantum extension of the thermodynamic uncertainty relation where dynamical fluctuations are quantified by the Terletsky--Margenau--Hill quasiprobability, a quantum generalization of the classical joint probability. 
    The obtained inequality plays a complementary role to existing quantum thermodynamic uncertainty relations, focusing on observables' change rather than exchange of charges through jumps and respecting initial coherence. 
    Quasiprobabilities show anomalous behaviors that are forbidden in classical systems, such as negativity; we reveal that negativity or a non-classically enhanced escape rate is necessary to increase an output-to-dissipation ratio beyond classical limitations and show that the requirements are basis-independent and stronger than quantum coherence. 
    To illustrate these statements, we employ a model that can exhibit a dissipationless heat current, which would be prohibited in classical systems; we construct a state that has much coherence but does not lead to a dissipationless current due to the absence of anomalous behaviors in quasiprobabilities. 
\end{abstract}

\maketitle

\noindent\textit{Introduction}.---
Finding universal lower bounds on the entropy production, which quantifies the irreversibility in the process, is one of the main tasks in nonequilibrium thermodynamics because they provide fundamental limitations beyond the second law of thermodynamics~\cite{shiraishi2023introduction}.
The thermodynamic uncertainty relation (TUR) is arguably the most crucial example, having been intensively studied over the last decade~\cite{barato2015thermodynamic,horowitz2020thermodynamic}. 
It consists of dissipation, quantified by the entropy production rate (EPR) $\epr$, and a general current's strength $J_X$ and fluctuations $S_X$, typically given in the form 
\begin{align}
    \epr\geq \frac{2J_X^2}{S_X}.
\end{align}
Incorporating the information of fluctuations explicitly, it provides a universal finite lower bound that tightens the second law, $\epr\geq 0$. 

While extending the TUR to the quantum regime is a crucial problem gathering much attention recently~\cite{agarwalla2018assessing,ptaszynski2018coherence,carollo2019unraveling,hasegawa2020quantum,van2022thermodynamics,hasegawa2023unifying,prech2025role,van2025fundamental,moreira2025precision,brandner2025thermodynamic,guarnieri2019thermodynamics,timpanaro2019thermodynamic,sacchi2021thermodynamic,miller2021thermodynamic,das2023precision,farina2024thermodynamic,hasegawa2021thermodynamic}, there is a fundamental issue of how to evaluate \textit{dynamical}  fluctuations. 
TURs usually involve fluctuations in fluxes or observables' change~\cite{gingrich2016dissipation,pietzonka2016universal,proesmans2017discrete,pietzonka2017finite,dechant2018multidimensional,hasegawa2019fluctuation,liu2020thermodynamic,falasco2020unifying,koyuk2020thermodynamic,vo2020unified,dechant2020fluctuation,yoshimura2021thermodynamic,lee2021universal,lee2022speed,delvenne2024thermokinetic,kwon2024unified,kolchinsky2024generalized,vo2025inverse,nagayama2025geometric}, hence they require statistics more than single time points. 
However, in quantum mechanics, physical quantities at different times do not commute in the Heisenberg sense, which makes fluctuations elusive. 

Conventionally, there are two approaches trying to characterize quantum fluctuations: one is based on the full counting statistics or continuous measurement~\cite{agarwalla2018assessing,ptaszynski2018coherence,carollo2019unraveling,hasegawa2020quantum,van2022thermodynamics,hasegawa2023unifying,prech2025role,van2025fundamental,moreira2025precision,brandner2025thermodynamic,guarnieri2019thermodynamics}, which identifies jumps as exchanging charges with the environment or detections of a signal in an experiment~\cite{landi2024current}. 
However, it cannot explore fluctuations of more general observables that are not directly related to the jumps. 
The other approach utilizes the two-point measurement~\cite{sacchi2021thermodynamic,miller2021thermodynamic,das2023precision,farina2024thermodynamic}, which, however, ends up with discarding coherence regarding the measured observable we are interested in, because it includes invasive measurement steps~\cite{esposito2009nonequilibrium}. 
Hence, these methods can fail to capture all the quantum effects on thermodynamic trade-off relations. 

Our idea to overcome this difficulty is to focus on \textit{quasiprobabilities} to fully describe the quantum statistics of observables that are not detected by jumps and may exhibit non-commutativity. 
Quasiprobabilities are quantum extensions of the classical joint probabilities, and their peculiar behaviors have been attracting significant attention in recent years~\cite{gherardini2024quasiprobabilities,lostaglio2023kirkwood,arvidsson2024properties}. 
While the classical probabilities satisfy positivity and linearity and lead to correct marginals, any quantum counterparts are known to violate at least one of these three properties~\cite{perarnau2017no,lostaglio2023kirkwood}. 
As the violation of positivity is connected to genuine quantumness called contextuality~\cite{spekkens2008negativity,lostaglio2018quantum}, quasiprobabilities that may take negative values have recently been studied extensively~\cite{ferrie2011quasi,allahverdyan2014nonequilibrium,yunger2018quasiprobability,gonzalez2019out,levy2020quasiprobability,hernandez2024projective,donelli2025orthogonalization}. 

\begin{figure}
    \centering
    \includegraphics[width=0.85\linewidth]{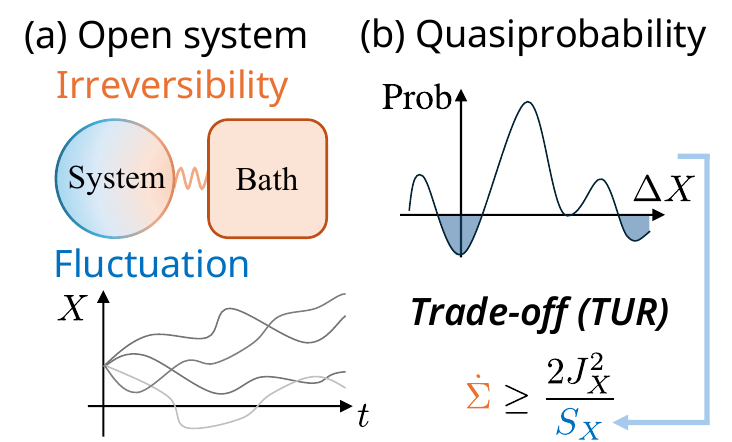}
    \caption{(a) In open systems, we cannot reduce two costs simultaneously: irreversibility, quantified by the entropy production rate $\epr$, and fluctuations. That is represented by a universal trade-off relation called the thermodynamic uncertainty relation (TUR), where the product between $\epr$ and the dynamical fluctuation $S_X$ of a physical quantity $X$ is bounded by a current strength $J_X$. (b) We prove a quantum TUR with $S_X=m_X$ [Eq.~\eqref{eq:qtur}], where the observable's dynamical fluctuation is quantified by the quasiprobability, a quantum extension of the classical joint probability that may take negative values. }
    \label{fig:fig0}
\end{figure}

In this Letter, we derive a quantum TUR by quantifying the dynamical fluctuation of an observable with such a quasiprobability for the first time (see Fig.~\ref{fig:fig0}). 
We consider Markovian open quantum systems and show that the short-time variance of an observable's change, as assessed by the Terletsky--Margenau--Hill quasiprobability~\cite{terletsky1937,margenau1961correlation}, provides a short-time TUR bound on the EPR for any state and observable. 
Moreover, on the basis of the non-classicality of quasiprobability, we elucidate criteria required for the anomalously large fluctuation absent in classical systems. 
Combined with our TUR, the criteria show that a large amount of heat flux with finite dissipation as reported in Refs.~\cite{tajima2021superconducting,funo2025symmetry} can happen only when the quasiprobability exhibits negative values or a non-classically enhanced escape rate. We further compare our quasiprobability-based conditions with the coherence-based ones discussed in Ref.~\cite{tajima2021superconducting}.

\noindent\textit{Setup}.---
We consider completely positive Markovian quantum dynamics, which are generally described by the Gorini--Kossakowski--Sudarshan--Lindblad (GKSL) equation~\cite{lindblad1976generators,gorini1976completely,breuer2002theory}
\begin{align}
    &\frac{d\rho}{dt}=\mathcal{L}(\rho)
    \coloneqq-i[H,\rho]+\diss(\rho),\label{eq:QME}\\
    &\diss(\rho)\coloneqq\sum_{k}L_k\rho L_k^\dagger -\frac{1}{2}\{L_k^\dagger L_k,\rho\}, 
\end{align}
where $\rho$ is the density operator, $H$ the Hamiltonian, $L_k$ jump operators, $[\cdot,\cdot]$ the commutator, and $\{\cdot,\cdot\}$ the anticommutator. 
We refer to $\mathcal{L}$ and $\diss$ as the Liouvillian and the dissipator, respectively. 
While we assume the Hamiltonian and jump operators are time-independent, our results are easily extended to time-dependent cases.
Given the initial state $\rho_0$, Eq.~\eqref{eq:QME} is formally solved by $\rho(t)=e^{\mathcal{L}t}\rho_0$. 
We indicate the adjoint of $\mathcal{L}$ and $\diss$ by $\liou^\dagger $ and $\diss^\dagger $. 
For an observable (Hermitian operator) $X$, we describe the eigendecomposition as $X=\sum_x x\Pi_x$ with eigenvalues $x$ and projectors $\Pi_x$. 
While we focus on time-independent observables for simplicity, our results are valid for time-dependent ones (detailed later).
Its expectation value at time $t$ is given by $\langle X\rangle_t =\tr(X\rho(t))=\sum_x x p(x,t)$, where $p(x,t)=\tr(\Pi_x\rho(t))$. 
We divide the time derivative into the Hamiltonian part $J_X^\mathrm{H}(t)=i\langle[H,X]\rangle_t$ and the dissipative part $J_X^\mathrm{d}(t)=\langle \diss^\dagger(X)\rangle_t$ as $d_t\langle X\rangle_t=J_X^\mathrm{H}(t)+J_X^\mathrm{d}(t)$. 

To discuss multi-time statistics of $X$ in the quantum dynamics beyond the single-time one $p(x,t)$, we introduce the quasiprobability~\cite{gherardini2024quasiprobabilities,lostaglio2023kirkwood,arvidsson2024properties}. 
In particular, we focus on the two-time statistics given by the Terletsky--Margenau--Hill (TMH) quasiprobability~\cite{terletsky1937,margenau1961correlation,pei2023exploring}, which is defined by 
\begin{align}
    q(y,t+\Delta t;x,t)
    \coloneq \frac{1}{2}\tr\Big(\big\{e^{\mathcal{L}^\dagger  \Delta t}\Pi_y,\Pi_x\big\}\rho(t)\Big). 
\end{align}
The TMH quasiprobability is the real part of the Kirkwood--Dirac quasiprobability~\cite{kirkwood1933quantum,dirac1945analogy}. While its marginals provide the single-time distributions as $\sum_{y}q(y,t+\Delta t;x,t)=p(x,t)$ and $\sum_{x}q(y,t+\Delta t;x,t)=p(y,t+\Delta t)$, it can become negative, unlike the classical joint probability distribution, which is an emergence of genuine quantumness~\cite{spekkens2008negativity,lostaglio2018quantum}. 

Analogously to the classical case~\cite{gardiner2009stochastic}, we define the moments of ``change in $X$'' by 
\begin{align}
    \big\langle (\Delta X)^n\big\rangle_{t,t+\Delta t}\coloneqq \sum_{x,y}(y-x)^nq(y,t+\Delta t;x,t). 
\end{align}
It can be computed from the moment generating function $\Gen_t(\lambda,\Delta t)$ as $\langle (\Delta X)^n\rangle_{t,t+\Delta t}=(-i\partial_\lambda)^n\Gen_t(\lambda,\Delta t)|_{\lambda=0}$, where $\Gen_t(\lambda,\Delta t)$ is defined as 
\begin{align}
    \Gen_t(\lambda,\Delta t)
    \coloneqq \frac{1}{2}\tr\Big(\big\{e^{\liou^\dagger \Delta t}e^{i\lambda X},e^{-i\lambda X}\big\}\rho(t)\Big). \label{eq:tmh_genfunc}
\end{align}
Theoretical studies have revealed that the moment generating function can be directly measured by devising an interferometric scheme~\cite{lostaglio2023kirkwood,gherardini2024quasiprobabilities}, which has been experimentally realized for $X=H$~\cite{hernandez2024interferometry}. 
Thus, $\langle (\Delta X)^n\rangle_{t,t+\Delta t}$ is not only theoretically but also practically a reasonable measure of the dynamical fluctuation in a quantum object $X$. 
Additionally, as shown in Supplemental Material~\cite{smArxiv}, $\Gen_t(\lambda,\Delta t)$ can be obtained by appropriately ``quantizing'' the classical counterpart. 

We further define the short-time moments $m_X^{(n)}(t)$ by $m_X^{(n)}(t)\coloneqq \lim_{\Delta t \to 0}\langle (\Delta X)^n\rangle_{t,t+\Delta t}/\Delta t$. 
We write $m_X^{(2)}(t)$ simply as $m_X(t)$ and call it the short-time fluctuation of $X$ because the case $n=2$ will be of the most importance due to its direct connection to the variance, $m_X^{(2)}(t)=\lim_{\Delta t\to0}\mathrm{Var}(\Delta X)/\Delta t$, where $\mathrm{Var}(\Delta X)\coloneqq \langle(\Delta X)^2\rangle_{t,t+\Delta t}-(\langle\Delta X\rangle_{t,t+\Delta t})^2$. 
The limit is always well defined because the TMH quasiprobability is expanded as $q(y,t+\Delta t;x,t)=\delta_{xy}p(x,t)+ T_{yx}\big(\rho(t)\big)\Delta t + O(\Delta  t^2)$
with
\begin{align}
    T_{yx}(\rho)\coloneqq \frac{1}{2}\tr\big(\{\mathcal{L}^\dagger  \Pi_y,\Pi_x\big\}\rho\big)
\end{align}
when $\Delta t$ goes to zero. 
Because $\liou^\dagger (I)=0$, we have $\sum_{y}T_{yx}(\rho)=0$. 
The short-time moments are given by $T_{yx}$ as  
\begin{align}
    m_X^{(n)}(t)=\sum_{x,y}(y-x)^n T_{yx}(\rho(t)). \label{eq:moments_with_T}
\end{align}
In the End Matter, we prove that Eq.~\eqref{eq:moments_with_T} is valid for $n=2$ even if $X$ depends on time.

We refer to $T_{yx}$ as the flux from $x$ to $y$ in analogy to the classical Markov processes. 
Indeed, if $\rho(t)$ commutes with $X$ and $X$ has no degeneracy, then $\rho$ is diagonalized as $\rho(t)=\sum_x p(x,t)\Pi_x$ with $\Pi_x=\ketbras{x}$ and we have $T_{yx}(\rho(t))=R_{yx}p(x,t)$ with $R_{yx}=\sum_k|\bra{y}L_k\ket{x}|^2$. 
Now, $p(x,t)$ and $R_{yx}$ can be interpreted respectively as the occupation probability of the classical state labeled by $x$ and the transition rate from state $x$ to state $y$~\cite{gardiner2009stochastic,esposito2006fluctuation}. 
The diagonal elements $T_{xx}(\rho(t))$ also read $R_{xx}p(x,t)$, where $R_{xx}=-\sum_{x'(\neq x)}\sum_{k}|\bra{x'}L_k\ket{x}|^2$, and $-R_{xx}$ corresponds to the escape rate from state $x$. 
Now that, even in the general (quantum) cases, we call the sum $\bar{\lambda}(\rho)\coloneqq -\sum_x T_{xx}(\rho)$ the average escape rate. 

Finally, we make a few assumptions to discuss thermodynamics. 
We assume that each jump $k$ has a unique counterpart $-k$ and that jump $k$ induces an entropy current $s_k$ such that $s_{-k}=-s_k$. The jump operators are postulated to satisfy the local detailed balance $L_k=e^{s_k/(2k_{\mathrm{B}})}L_{-k}^\dagger$~\cite{maes2021local,horowitz2013entropy,strasberg2022quantum}. 
Then, we can define the entropy production rate by 
\begin{align}
    \epr\big(\rho(t)\big)
    \coloneq k_{\mathrm{B}}\frac{d}{dt}S(t)+\sum_k s_k\tr\big(L_k^\dagger L_k\rho(t)\big), \label{eq:epr}
\end{align}
where $S(t)\coloneq -\tr(\rho(t)\ln\rho(t))$ is the von Neumann entropy. 
Hereafter, we set the Boltzmann constant $k_\mathrm{B}$ to one for simplicity. 

\noindent\textit{Main result}.---
The following inequality is our main result:
\begin{align}
    \epr\big(\rho(t)\big)\geq \frac{2|J_X^\mathrm{d}(t)|^2}{m_X(t)} \label{eq:qtur}
\end{align}
for any state $\rho(t)$ and observable $X$. 
The numerator is the dissipative part of the time derivative of $\langle X\rangle_t$, so it quantifies the changing rate of $X$ due to the dissipative dynamics. 
On the other hand, the denominator, the short-time fluctuation, represents the dynamical fluctuation of $X$ as evaluated by the TMH quasiprobability. 
Thus, the inequality represents the universal trade-off between dissipation and fluctuations, i.e., the TUR~\cite{barato2015thermodynamic,horowitz2020thermodynamic}. 
Specifically, it generalizes the so-called short-time TUR in classical systems~\cite{otsubo2020estimating} (also known as the entropic bound~\cite{dechant2018current,kwon2024unified}), which have been utilized in estimating the EPR~\cite{seifert2019stochastic,li2019quantifying,otsubo2020estimating,manikandan2020inferring,dechant2021improving} and deriving the power-efficiency trade-off~\cite{shiraishi2016universal,kwon2024unified}. 
This connection will become clear if we assume that $\rho$ commutes with $X$; then, the numerator and denominator read $J_X^\mathrm{d}(t)=d_t\langle X\rangle_t$ and $m_X(t)=\sum_{x,y}(x-y)^2R_{yx}p(x,t)$, which appear in the short-time TUR presented in Ref.~\cite{otsubo2020estimating}. 
As discussed in Ref.~\cite{kwon2024unified}, the classical short-time TUR (entropic bound) holds without modification when the system is under a time-dependent protocol or has odd variables~\footnote{Note that Ref.~\cite{kwon2024unified} presents the bound in the time-integrated form, but it is equivalent to the instantaneous form.}, unlike the finite-time versions~\cite{koyuk2020thermodynamic,lee2021universal}; our TUR also applies to time-dependent systems, and we expect that it holds even if the system's dynamics is not time-reversal symmetric.

Let us prove Eq.~\eqref{eq:qtur}. We employ the following inequality derived by one of the present authors in Ref.~\cite{yoshimura2025force}: 
\begin{align}
    \epr\big(\rho(t)\big)\geq \frac{|\tr(X\diss(\rho(t)))|^2}{\diff_X(\rho(t))}, \label{eq:tur_old}
\end{align}
where $\diff_X(\rho)=\frac{1}{2}\tr(\rho(\diss^\dagger (X^2)-\{\diss^\dagger (X),X\}))$ is called the quantum diffusivity (the derivation is reviewed in Supplemental Material~\cite{smArxiv}). 
It is easy to see $\tr(X\diss(\rho(t)))=J_X^\mathrm{d}(t)$. 
The nontrivial point is that we can also associate the denominator $\diff_X(\rho(t))$ with $m_X(t)$. First, by expanding the definition, we find
\begin{align}
    m_X(t) 
    &= \frac{1}{2}\sum_{x,y}(y-x)^2 \tr(\{\liou^\dagger \Pi_y,\Pi_x\}\rho(t))\\
    &= \tr\big(\liou^\dagger (X^2)\rho(t)\big)
    - \tr\big(\{\liou^\dagger (X),X\}\rho(t)\big). \label{eq:mxexpression}
\end{align}
Moreover, by using an identity $\{[H,X],X\}=[H,X^2]$, we can remove the Hamiltonian part from $\liou^\dagger$. 
Therefore, we find the equality 
\begin{align}
    \diff_X\big(\rho(t)\big)=\frac{1}{2}m_X(t). \label{eq:dequalm}
\end{align}
Combining it with Eq.~\eqref{eq:tur_old}, we obtain the TUR~\eqref{eq:qtur}. 
Note that this discussion applies to time-dependent observables since Eq.~\eqref{eq:moments_with_T} remains valid, as discussed in the End Matter. 

\noindent\textit{Comparison with existing TURs}.---We compare our TUR~\eqref{eq:qtur} with existing quantum TURs. 
First, we stress the difference from widely studied methods, the full counting statistics (FCS) approach~\cite{landi2024current} and the two-point measurement (TPM) approach~\cite{esposito2009nonequilibrium}. 
In short, our result is complementary to those previous ones. 
While the TURs based on the FCS~\cite{carollo2019unraveling,hasegawa2020quantum,van2022thermodynamics,hasegawa2023unifying,prech2025role,van2025fundamental,brandner2025thermodynamic,guarnieri2019thermodynamics,timpanaro2019thermodynamic} deal with current observables associated with the counting of jumps, our TUR is given by the fluctuations of intrinsic observables (i.e., Hermitian operators on the system's Hilbert space), which do not have to be connected to jumps. 
It is noteworthy that in classical Markov jump processes, an intrinsic observable's change is entirely given by keeping track of jumps, because every physical quantity takes determined values in classical states~\cite{gardiner2009stochastic}. 
However, quantum states linked by quantum jumps are not eigenstates of every intrinsic observable (they may not be even for the Hamiltonian; see Model A in \cite{landi2024current}). 
In the End Matter, we discuss a nontrivial intersection between the FCS and quasiprobabilities when $[X, L_k]=w_kL_k$ holds for an observable $X$ with certain weights $w_k$. 

The TPM method, adopted to quantify dynamical fluctuations in Refs.~\cite{sacchi2021thermodynamic,miller2021thermodynamic,das2023precision,farina2024thermodynamic}, inevitably leads to decoherence in the initial density operator due to invasive initial measurement, despite having a clear experimental perspective. 
Thus, derived TURs fail to capture initial coherence's effects.
On the other hand, in our TUR, every initial coherence is incorporated owing to the employment of the quasiprobability, while maintaining experimental access by the interferometric method~\cite{lostaglio2023kirkwood,gherardini2024quasiprobabilities,hernandez2024interferometry}. 

Finally, we mention the connection to the ``TUR'' [Eq.~\eqref{eq:tur_old}] given in Ref.~\cite{yoshimura2025force}. 
Despite the apparent similarity, our TUR has a substantial advantage over the previous result. 
As is clear from the proof, our TUR is derived by combining their result and the equality~\eqref{eq:dequalm}. 
The latter equality gives a clear statistical meaning to $\diff_X(\rho)$, which was interpreted as a fluctuation measure only in classical cases in Ref.~\cite{yoshimura2025force}. 
Moreover, as discussed below, the quasiprobabilistic perspective brought by Eq.~\eqref{eq:dequalm} enables us to understand a non-classical suppression of dissipation through non-classical behaviors of quasiprobabilities, which cannot be understood from Eq.~\eqref{eq:tur_old} alone. 

\noindent\textit{Anomalous scaling via non-classicality of quasiprobability}.---
When rewritten as 
\begin{align}
    Q_X(\rho)\coloneqq \frac{|J_X^{\mathrm{d}}(\rho)|^2}{\epr(\rho)}\leq \frac{1}{2}m_X(\rho), \label{eq:tur_rewrite}
\end{align}
the TUR shows that the output-to-dissipation ratio $Q_X(\rho)$ is upper bounded by the short-time fluctuation (hereafter, we write $J_X^\mathrm{d}(\rho)$ and $m_X(\rho)$ to indicate $J_X^{\mathrm{d}}(t)$ and $m_X(t)$ when $\rho(t)=\rho$). 
When $X$ is extensive, it is expected that $Q_X(\rho)$ can grow at most at the order of system size. 
Intriguingly, in Ref.~\cite{tajima2021superconducting}, the authors show that in a system with an $N$-fold degenerate Hamiltonian $H$, we can let $J_H^\mathrm{d}(\rho)$ grow at $O(N)$, keeping $\epr(\rho)=O(1)$, which means $Q_H(\rho)=O(N^2)$, violating the classical limit of $O(N)$.
They call such a phenomenon \textit{dissipationless current}.
They also discuss that for a certain class of degenerate eigenbases, the dissipationless current can be attributed to a large amount of quantum coherence in $\rho$.

Owing to the quasiprobability TUR~\eqref{eq:qtur}, which connects dissipation and quasiprobabilities, we can characterize this non-classical phenomenon in an eigenbasis-free way. 
We investigate the scaling of $m_X(\rho)$ for an $N$-fold degenerate observable $X$ instead of $Q_X(\rho)$ because its fast scaling is a necessary condition from Eq.~\eqref{eq:tur_rewrite}. 
To understand the effect of degeneracy on the scaling, we assume that the variety of $X$'s eigenvalue and its operator norm are $O(1)$ and that the degree of degeneracy is independent of the eigenvalue; thus, we assume the form $X=\sum_{s}\sum_{j=1}^Nx_s\Pi_{s,j}$ with $\Pi_{s,j}=\ketbras{s,j}$ and $X\ket{s,j}=x_s\ket{s,j}$. This includes the Hamiltonian considered in Ref.~\cite{tajima2021superconducting}. 
We call the $O(N^\alpha)$ scaling of $m_X(\rho)$ with $\alpha>1$ \textit{anomalous scaling}.

Before stating the result, we introduce a couple of quantities.
For an eigenbasis, the flux from $(s,j)$ to $(s',j')$ and the average escape rate are respectively defined as $T_{s'j'sj}(\rho)=\frac{1}{2}\tr(\{\liou^\dagger \Pi_{s',j'},\Pi_{s,j}\}\rho)$ and $\bar{\lambda}(\rho)=-\sum_{s,j}T_{sjsj}(\rho)$. 
We further define the integrated fluxes $\mathcal{T}_{s's}(\rho)$ as 
\begin{align}
    \mathcal{T}_{s's}(\rho)
    &\coloneqq
    \begin{cases}
        \sum_{j,j'}T_{s'j'sj}(\rho)&(s\neq s')\\
        \sum_{j,j'(\neq j)}T_{sj'sj}(\rho)&(s= s')
    \end{cases}.
\end{align}
The equality $\sum_{s,s'}\mathcal{T}_{s's}(\rho)-\bar{\lambda}(\rho)=0$ follows from the property $\sum_{s',j'}T_{s'j'sj}(\rho)=0$. 
The short-time fluctuation is given solely by the integrated fluxes as $m_X(\rho)=\sum_{s,s'}(x_s-x_{s'})^2\mathcal{T}_{s's}(\rho)$. 

The next statement is our second main result: \textit{if neither of the following conditions is satisfied for some eigenbasis, $m_X(\rho)$ does not exhibit anomalous scaling}: 
\begin{align*}
    \text{\textit{(Q1)}\;\;}&\exists s,s',\;\;
    \lim_{N\to\infty}\frac{\mathcal{T}_{s's}(\rho)}{N}=-\infty,\\
    \text{\textit{(Q2)}\;\;}& \lim_{N\to\infty}\frac{\bar{\lambda}(\rho)}{N}=\infty. 
\end{align*}
The first condition (Q1) represents significant negativity in the fluxes, which results in negative quasiprobability because then $T_{s'j'sj}(\rho)<0$ for several $j,j'$ and 
\begin{align}
    &q\big((s',j'),t+\Delta t;(s,j),t\big)\notag\\
    &= T_{s'j'sj}(\rho(t))\Delta t +O(\Delta t^2)<0. 
\end{align}
Note that the negativity of the quasiprobability also indicates the genuine quantumness called contextuality~\cite{spekkens2008negativity,lostaglio2018quantum}. 
The other one (Q2) is also non-classical because it means the chance of escape grows more rapidly than the number of evacuation targets increases~\footnote{In other words, if we adopt conditions (1) $O(1)$ transition rates and (2) $O(1)$ channel multiplicity, which are detailed two paragraphs later, as the definition of classicality, the escape rate scales at most $O(N)$ in classical situations.}. 
The statement shows that either of these quantumness conditions has to hold in \textit{any} basis for the anomalous scaling to occur. 

The necessity of either (Q1) or (Q2) for anomalous scaling is proved by showing that it is equivalent to the existence of a pair $(s,s')$ such that $|\mathcal{T}_{s's}(\rho)|/N\to\infty$ as $N\to\infty$ (for the proof, see the End Matter). Under this equivalence, it becomes evident that if the conditions (Q1) and (Q2) are violated, we do not observe anomalous scaling since $m_X(\rho)\leq \sum_{s,s'}(x_s-x_{s'})^2|\mathcal{T}_{s's}|$. We stress that the split of the formal condition into (Q1) and (Q2) enables us to bridge anomalous scaling and the nonclassicality of quasiprobabilities.

When $\rho$ commutes with $X$, the quasiprobability conditions necessitate non-classicality at the level of basis components for every simultaneous eigenbasis. 
Now, we define an eigenbasis $\{\ket{s,j}\}$ to be classical if the following conditions are satisfied for all $(s,j)$ and $(s',j')(\neq(s,j))$: as $N$ increases, (1) the jump rate does not grow ($|\bra{s',j'}L_k\ket{s,j}|=O(1)$), and (2) the number of jumps that mediate the states does not grow ($\#\{k\colon|\bra{s',j'}L_k\ket{s,j}|>0\}=O(1)$)~\footnote{These conditions on bare transition rates will be naturally satisfied if we consider few-body systems or systems with only short-range interactions.}. 
Then, if a simultaneous eigenbasis $\{\ket{s,j}\}$ of $\rho$ and $X$ is classical, we can prove that the anomalous scaling does not occur. 
First observe that $-R_{sjsj}=\sum_{s',j'(\neq s,j)}\sum_k |\bra{s',j'}L_k\ket{s,j}|^2=O(N)$, and thus $\bar{\lambda}(\rho)=-\sum_{s,j}R_{sjsj}p_{sj}=O(N)$, where $p_{s,j}=\tr(\rho\Pi_{s,j})$. 
Besides, we have $T_{s'j'sj}(\rho)=\sum_k |\bra{s',j'}L_k\ket{s,j}|^2p_{s,j}\geq 0$ since $\rho$ is diagonal with respect to $\{\ket{s,j}\}$. 
Thus, $m_X(\rho)$ can grow at most at $O(N)$ because both (Q1) and (Q2) are violated by the eigenbasis.  
This is in accordance with the result of Ref.~\cite{tajima2021superconducting}, where dissipationless current is proved not to occur when $\rho$ has no coherence with respect to this type of eigenbasis.

\begin{figure}
    \centering
    \includegraphics[width=\linewidth]{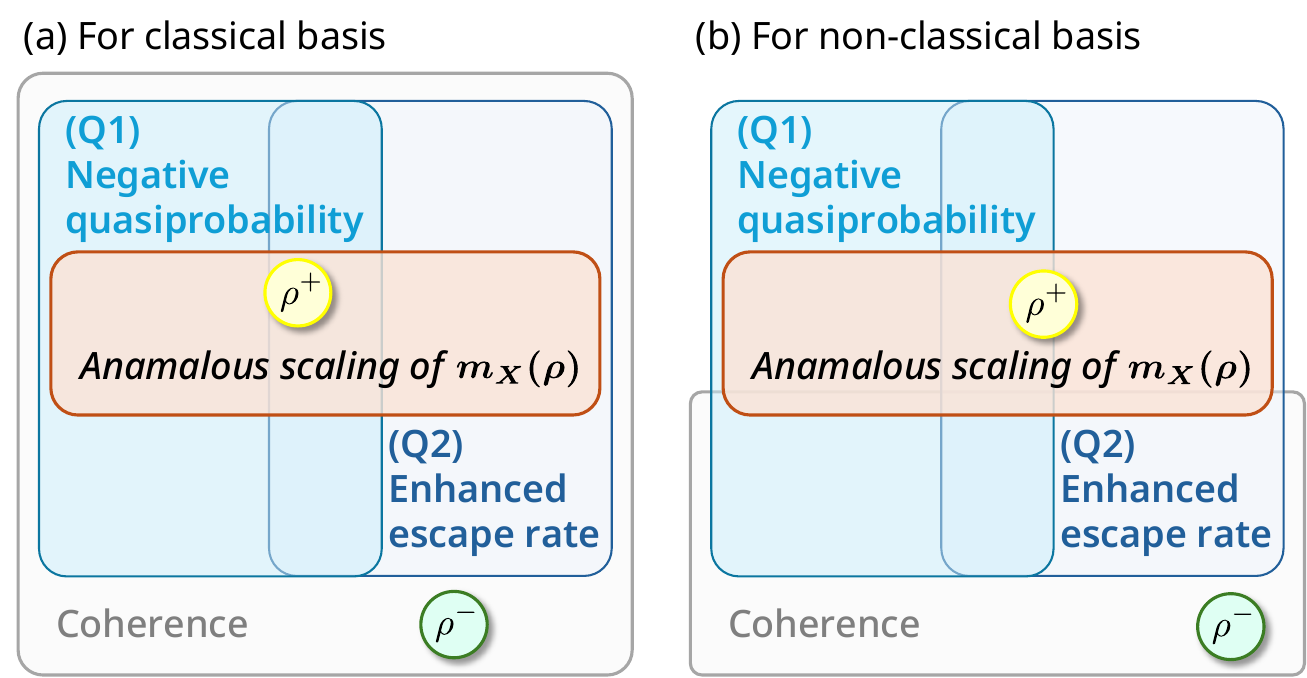}
    \caption{Hierarchy of conditions.
    Condition (Q1) or (Q2) is always required for $m_X(\rho)$ to scale anomalously, which is necessary for the output-to-dissipation ratio $Q_X(\rho)$ to grow faster than the classical limit. 
    (a) For classical eigenbases, our necessary conditions are tighter than the condition that the state has more than $O(1)$ coherence.
    (b) If we take a non-classical eigenbasis, coherence regarding that basis loses its connection to the scaling. In contrast, our two conditions remain relevant. 
    In the example, we examine two characteristic cases, $\rho^+$ and $\rho^-$.
    }
    \label{fig:fig1}
\end{figure}

In addition, we can show that neither (Q1) nor (Q2) is realized unless $\rho$ has significant (more than $O(1)$) coherence~\cite{smArxiv}; hence, our quasiprobability-based conditions are tighter than having divergent coherence, as depicted in Fig.~\ref{fig:fig1}(a). 
Moreover, given at the quasiprobability level, the criteria apply to any eigenbasis (see Fig.~\ref{fig:fig1}(b))~\footnote{When a non-classical eigenbasis is used, the corresponding classical Markov jump process has transition rates that scale as $O(N^\alpha)$ with some positive exponent $\alpha$. 
However, this scaling is typically not possible in realistic classical systems~\cite{shiraishi2016universal}, and therefore the classicality of the eigenbasis is implicitly assumed in Ref.~\cite{tajima2021superconducting}.}. 
The relationship is illustrated later through an example.

\noindent\textit{Example}.---
We illustrate the above discussion through the model used in Ref.~\cite{tajima2021superconducting}. 
Its Hamiltonian $H$ has two $N$-fold degenerate eigenvalues, $0$ and $\omega$.
The jump operators represent simultaneous jumps between energy levels; with an eigenbasis $\{\ket{g,j},\ket{e,j}\}_{j=1}^N$ such that $H\ket{g,j}=0$ and $H\ket{e,j}=\omega\ket{e,j}$, they are defined as $L_+=\sqrt{\gamma_+}\sum_{j,j'}\ketbrad{e,j}{g,j'}$ and $L_-=\sqrt{\gamma_-}\sum_{j,j'}\ketbrad{g,j}{e,j'}$. 
Anomalous scaling $m_H(\rho)=O(N^2)$ is observed if we set $\rho=\rho^+=p_g\ketbras{g,+}+p_e\ketbras{e,+}$ with $\ket{s,+}=\frac{1}{\sqrt{N}}\sum_{j}\ket{s,j}$ for $s=e,g$~\cite{smArxiv}. 
The state $\rho^+$ has a large $l^1$-coherence $C_{l^1}(\rho^+)=O(N)$, defined as $C_{l^1}(\rho)=\sum_{s,j,s',j'|(s,j)\neq (s',j')}|\bra{s,j}\rho\ket{s',j'}|$, to which Ref.~\cite{tajima2021superconducting} attributed the dissipationless current. 
Now, note that the basis $\{\ket{g,j},\ket{e,j}\}_{j=1}^N$, defining the $l^1$-coherence, is classical since there are only two jumps ($k=\pm$) regardless of $N$ and, for example, $\bra{e,j'}L_+\ket{g,j}=\sqrt{\gamma_+}=O(1)$. 

The hierarchy between conditions is exemplified by a state $\rho^-$ that has as much coherence regarding the classical eigenbasis as $\rho^+$ but does not yield dissipationless current (see Fig.~\ref{fig:fig1}(a)). 
Such a state is constructed by replacing $\ket{s,+}$ in the definition of $\rho^+$ with $\ket{s,-}=\frac{1}{\sqrt{N}}\sum_{j}(-1)^j\ket{s,j}$. 
For this state, the integrated fluxes and average escape rate read $\mathcal{T}_{s's}(\rho^-)=\gamma_sp_s\chi(N)$ for $s\neq s'$, $\mathcal{T}_{ss}(\rho^-)=\frac{\gamma_sp_s}{2}(N-\chi(N))$, and $\bar{\lambda}(\rho^-)=\frac{\gamma_+p_g+\gamma_-p_e}{2}(N+\chi(N))$, where $\gamma_g=\gamma_+$, $\gamma_e=\gamma_-$, and $\chi(N)$ is the remainder when $N$ is divided by $2$~\cite{smArxiv}. 
Thus, $\rho^-$ does not satisfy either (Q1) and (Q2), and leads to $m_H(\rho^-)=O(1)$. 
This example illustrates that the non-classicality represented via the quasiprobability is more restrictive than requiring a large amount of coherence, although both are just necessary conditions.

Finally, we remark that if we consider a non-classical basis including $\ket{g,+}$ and $\ket{e,+}$ (note that $\bra{e,+}L_+\ket{g,+}=\sqrt{\gamma_+}N$), the state $\rho^+$ has no coherence regarding this basis. 
This fact shows that $m_H(\rho^+)$ may exhibit anomalous scaling (that is, we may have dissipationless current) even if $\rho^+$ has no coherence if the referred basis is not classical (see Fig.~\ref{fig:fig1}(b)). 
Nonetheless, we see that $\rho^+$ fulfills either (Q1) or (Q2) for any eigenbasis, from the general discussion.  

\noindent\textit{Discussion}.---
We have explored fundamental connections between quantum thermodynamics and quasiprobability; in particular, we have derived a quantum TUR evaluating fluctuations through the TMH quasiprobability. 
We have also demonstrated that when discussing anomalous scaling of fluctuations in the TUR, which brings reduction of dissipation beyond classical limitations, non-classical behaviors of quasiprobability are more essential than quantum coherence. 
Although our results are derived based on the GKSL equation, which requires several conditions such as weak coupling between the system and the bath~\cite{breuer2002theory}, the quasiprobability can be defined regardless of the dynamics; hence, we expect that the results can be insightful beyond the GKSL regime.

To the best of our knowledge, our study is the first to bridge quasiprobabilities and the thermodynamic trade-off relations (except for a recent attempt~\cite{chun2025power} in a restricted situation, where negativity never appears). 
In addition to the TUR, several fundamental relations have been unveiled in the field, such as thermodynamic speed limits~\cite{shiraishi2018speed,funo2019speed,nakazato2021geometrical,van2021geometrical,hamazaki2022speed,yoshimura2023housekeeping,van2023thermodynamic,hamazaki2024speed,hamazaki2024quantum,nagayama2025infinite,ikeda2025speed}, bounds on asymmetry~\cite{ohga2023thermodynamic,liang2023thermodynamic,van2024dissipation,liang2024thermodynamic}, and fluctuation-response inequalities~\cite{ptaszynski2024dissipation,kwon2025fluctuation,aslyamov2025nonequilibrium}. 
It would be intriguing to study how quasiprobabilities will help explore non-classical features in their quantum extensions. 

\begin{acknowledgments}
    \noindent\textit{Acknowledgments.}---
    The authors are grateful to Hiroyasu Tajima, Ken Funo, and Yuuya Chiba for fruitful discussions. 
    K.Y.\ is supported by the Special Postdoctoral Researchers Program at RIKEN and JSPS KAKENHI Grant No.~24H00834. 
    R.H.\ is supported by JSPS KAKENHI Grant No.~JP24K16982.
    K.Y.\ and R.H.\ are supported by JST ERATO Grant No.~JPMJER2302, Japan.
\end{acknowledgments}

\input{biblio.bbl}

\section{End Matter}

\subsection{Validity of Eq.~\eqref{eq:moments_with_T} for time-dependent observables}
Assume that the observable of interest $X(t)$ depends on time and is decomposed as 
\begin{align}
    X(t)=\sum_s x_s(t) \Pi_s(t), 
\end{align}
where $\Pi_s(t)=\ketbras{x_s(t)}$. Here, $X$ may degenerate. 
We assume that $x_s(t)$ and $\Pi_s(t)$ smoothly change in time. 
Then, we can define the quasiprobability and the moments as 
\begin{align}
    &q(s',t+\Delta t;s,t)\notag\\
    &\quad\coloneqq \frac{1}{2}\tr\Big(\big\{e^{\mathcal{L}^\dagger  \Delta t}(\Pi_{s'}(t+\Delta t)),\Pi_s(t)\big\}\rho(t)\Big),\\
    &\big\langle (\Delta X)^n\big\rangle_{t,t+\Delta t}\notag\\
    &\quad\coloneqq \sum_{s,s'}\big(x_{s'}(t+\Delta t)-x_s(t)\big)^nq(s',t+\Delta t;s,t). 
\end{align}
Therefore, the short-time moments include contributions from the explicit time dependence of $X(t)$ as 
\begin{align}
    m_X^{(n)}(t)
    = &\sum_{s,s'}n(x_{s'}-x_s)^{n-1}\dot{x}_{s'}\delta_{ss'}\tr(\Pi_s\rho)\notag\\
    &+\sum_{s,s'}(x_{s'}-x_s)^{n}
    T_{x_{s'}x_s}(\rho)\notag\\
    &+\frac{1}{2}\sum_{s,s'}(x_{s'}-x_s)^n
    \tr\big(\{\dot{\Pi}_{s'},\Pi_s\}\rho\big), \label{eq:short_time_moments_time_dependent}
\end{align}
where the $t$ dependence is made implicit and a dot indicates time derivative as $\dot{x}_s=\frac{d}{dt}x_s(t)$ and $\dot{\Pi}_s=\frac{d}{dt}\Pi_s(t)$. 
While the second term in the right-hand side yields the time-independent result~\eqref{eq:moments_with_T}, additional terms arise from $X(t)$'s time dependence. 
Nonetheless, we can prove for even $n$
\begin{align}
    m_X^{(n)}(t)
    = \sum_{s,s'}(x_{s'}-x_s)^{n}
    T_{x_{s'}x_s}(\rho) \label{eq:even_short_time}
\end{align}
even if $X$ depends on time. 
Therefore, $\frac{1}{2}m_X(t)=\diff_X(\rho)$ remains valid for time-dependent observables (note that $\diff_X(\rho)$ is defined regardless of $X(t)$'s time dependence). 

To prove Eq.~\eqref{eq:even_short_time}, we first show the equality 
\begin{align}
    \{\dot{\Pi}_{s'},\Pi_s\}
    +\{\dot{\Pi}_{s},\Pi_{s'}\}
    =2\delta_{ss'}\dot{\Pi}_{s}. \label{eq:pidot_equality}
\end{align}
To this end, let us confirm the basic property of the projector, 
\begin{align}
    \Pi_{s'}(t+\Delta t)\Pi_s(t+\Delta t)=\delta_{ss'}\Pi_{s}(t+\Delta t). \label{eq:projector_deltat}
\end{align}
Plugging the expansion $\Pi_s(t+\Delta t)=\Pi_s(t)+\dot{\Pi}_s(t)\Delta t+O(\Delta t^2)$ into Eq.~\eqref{eq:projector_deltat} and comparing $O(\Delta t)$ terms, we find 
\begin{align}
    \dot{\Pi}_{s'}(t)\Pi_s(t)
    + \Pi_{s'}(t)\dot{\Pi}_s(t)
    = \delta_{ss'}\dot{\Pi}_s(t). 
\end{align}
Because this is valid under the swap $s\leftrightarrow s'$, we obtain Eq.~\eqref{eq:pidot_equality}. 

Now, it is easy to see that the third term of Eq.~\eqref{eq:short_time_moments_time_dependent} vanishes when $n$ is even. 
This is because then $(x_{s'}-x_s)^n$ becomes invariant under the swap of $s$ and $s'$, and thus 
\begin{align}
    &\frac{1}{2}\sum_{s,s'}(x_{s'}-x_s)^n
    \tr\big(\{\dot{\Pi}_{s'},\Pi_s\}\rho\big)\notag\\
    &= \frac{1}{4}\sum_{s,s'}(x_{s'}-x_s)^n
    \tr\Big[\big(\{\dot{\Pi}_{s'},\Pi_s\}+\{\dot{\Pi}_{s},\Pi_{s'}\}\big)\rho\Big]\\
    &=\frac{1}{2}\sum_{s,s'}(x_{s'}-x_s)^n\delta_{ss'}\tr(\dot{\Pi}_s\rho)
    =0. 
\end{align}
In addition, because the first term in Eq.~\eqref{eq:short_time_moments_time_dependent} automatically vanishes if $n>1$, we have Eq.~\eqref{eq:even_short_time}.

\subsection{Proof of equivalence between conditions}
Let us prove the equivalence between ``(Q1) or (Q2)'' and the existence of a pair $(s,s')$ such that $|\mathcal{T}_{s's}(\rho)|/N\to\infty$ as $N\to\infty$. 
We indicate the latter condition as (QA). 
It is obvious that (Q1) and (Q2) are individually sufficient to prove (QA) because $\bar{\lambda}(\rho)=\sum_{s,s'}\mathcal{T}_{s's}(\rho)$. 
On the other hand, the negation of (Q2) implies that $\sum_{s,s'}\mathcal{T}_{s's}(\rho)$ is $O(N)$. 
Then, we can split the summation as $\sum_{s,s'}\mathcal{T}_{s's}(\rho)= \mathcal{T}^+-\mathcal{T}^-$ with $\mathcal{T}^+=\sum_{s,s'|\mathcal{T}_{s's}(\rho)\geq 0}\mathcal{T}_{s's}(\rho)$ and $\mathcal{T}^-=\sum_{s,s'|\mathcal{T}_{s's}(\rho)<0}|\mathcal{T}_{s's}(\rho)|$. 
Due to the negation of (Q1), the negative part $\mathcal{T}^-$ is at most $O(N)$. Therefore, the positive part $\mathcal{T}^+$ can also be at most $O(N)$. This means the negation of (QA) and completes the proof.

\subsection{Connection between FCS and quasiprobability}
We discuss a nontrivial intersection between the FCS and quasiprobabilities. 
In the former framework, we consider current observables $\mathcal{J}_w\coloneqq \sum_k w_k \mathcal{N}_k$ during time interval $[t,t+\Delta t]$, where $w_k\in\mathbb{R}$ is the weight and $\mathcal{N}_k$ is the number of occurrences of jump $k$ during the time interval.
Its moments are gained from the moment generating function $\Gen_t^{\fcs}(\lambda,\Delta t)=\tr(e^{\liou_{\lambda}\Delta t}\rho(t))$ with $\liou_\lambda(\rho)=\liou(\rho)+\sum_k(e^{i\lambda w_k}-1)L_k\rho L_k^\dagger$~\cite{landi2024current}. 
As explained in the main text, the moments of $\Delta X$ also have their own generating function $\Gen_t(\lambda,\Delta t)$ defined in Eq.~\eqref{eq:tmh_genfunc}. 

While $\Gen_t$ and $\Gen_t^{\fcs}$ are different quantities, we can relate them when the jumps induce changes in $X$ as
\begin{align}
    L_k = \sum_{x,y|y-x=w_k} a_{yx} \ketbrad{y}{x}
\end{align}
with some $a_{yx}\in\mathbb{C}$. 
This condition is equivalent to
\begin{align}
    [X, L_k] = w_k L_k, \label{eq:commutation}
\end{align}
and then $\mathcal{J}_w$ can be regarded as accumulating the change of $X$. 
In this case, we can show 
\begin{align}
    \gen_t^{\fcs}(\lambda)
    -\gen_t(\lambda)
    = \sum_{x,y}H_{yx}\rho_{xy}\sin\big(\lambda(y-x)\big), \label{eq:g_difference}
\end{align}
where $\gen_t(\lambda)=\partial_{\Delta t}\Gen_t(\lambda,\Delta t)|_{\Delta t=0}$, $\gen_t^\fcs(\lambda)=\partial_{\Delta t}\Gen_t^\fcs(\lambda,\Delta t)|_{\Delta t=0}$, $H_{yx}=\bra{y}H\ket{x}$, and $\rho_{xy}=\bra{x}\rho(t)\ket{y}$. 
The functions $\gen_t(\lambda)$ and $\gen_t^\fcs(\lambda)$ provide the short-time moments because 
\begin{align}
    (-i\partial_\lambda)^n\gen_t(\lambda)|_{\lambda=0}
    &= \partial_{\Delta t}\langle (\Delta X)^n\rangle_{t,t+\Delta t}|_{\Delta t = 0}\\
    &= \lim_{\Delta t \to 0}\frac{\langle (\Delta X)^n\rangle_{t,t+\Delta t}}{\Delta t}, 
\end{align}
and the same discussion is possible for $\gen_t^\fcs(\lambda)$. 
From Eq.~\eqref{eq:g_difference}, we see that the counting statistics can provide the short-time statistics of $\Delta X$ if Eq.~\eqref{eq:commutation} holds and $H$ or $\rho$ commutes with $X$. 
If they do not, the unitary time evolution gives rise to changes in $X$, which cannot be detected through jumps. 
Nonetheless, even in the presence of such noncommutativity, the even-order moments coincide; 
\begin{align*}
    &(-i\partial_\lambda)^{2n}\gen_t(\lambda)|_{\lambda=0}
    -(-i\partial_\lambda)^{2n}\gen_t^{\fcs}(\lambda)|_{\lambda=0}\\
    &=(-1)^n\sum_{x,y}(y-x)^{2n}\sin\big(\lambda(y-x)\big)|_{\lambda=0}=0.
\end{align*}
Consequently, $m_X$ can be computed by monitoring jumps under  Eq.~\eqref{eq:commutation}. 
Still, we stress that Eq.~\eqref{eq:commutation} is a highly restrictive condition, so that neither of the TURs implies the other in general. 

Let us prove Eq.~\eqref{eq:g_difference}. It is not difficult to see that Eq.~\eqref{eq:commutation} leads to the following relations: 
\begin{subequations}
    \begin{gather}
        [X,L_k^\dagger]=-w_kL_k^\dagger, \label{eq:conseqA}\\
        [X,L_k^\dagger L_k]=0, \label{eq:conseqB}\\
        e^{i\lambda w_k}L_k=e^{i\lambda X}L_k e^{-i\lambda X}, \label{eq:conseqC}\\
        e^{i\lambda w_k}L_{k}^\dagger=e^{-i\lambda X}L_k^\dagger e^{i\lambda X}. \label{eq:conseqD}
    \end{gather}
\end{subequations}
Next, we can write $\gen_t^{\fcs}(\lambda)$ and $\gen_t(\lambda)$ as 
\begin{align}
    &\gen_t^{\fcs}(\lambda)
    =\tr(\liou_\lambda\rho(t))
    =\sum_k(e^{i\lambda w_k}-1)\tr(L_k^\dagger L_k\rho)\\
    &\gen_t(\lambda) 
    =\frac{i}{2}\tr\Big(\big\{[H,e^{i\lambda X}],e^{-i\lambda X}\}\rho(t)\Big)\notag \\
    &+\frac{1}{2}\sum_k\tr\Big(\big\{L_k^\dagger e^{i\lambda X}L_k-\frac{1}{2}\{L_k^\dagger L_k,e^{i\lambda X}\},e^{-i\lambda X}\big\}\rho(t)\Big). 
\end{align}
The summand in $\gen_t(\lambda)$ is transformed as 
\begin{align*}
    &\big\{L_k^\dagger e^{i\lambda X}L_k-\frac{1}{2}\{L_k^\dagger L_k,e^{i\lambda X}\},e^{-i\lambda X}\big\}\\
    &=L_k^\dagger e^{i\lambda X}L_ke^{-i\lambda X}
    +e^{-i\lambda X}L_k^\dagger e^{i\lambda X}L_k\\
    &\quad -L_k^\dagger L_k
    - \frac{1}{2}e^{i\lambda X}L_k^\dagger L_ke^{-i\lambda X}
    - \frac{1}{2}e^{-i\lambda X}L_k^\dagger L_ke^{i\lambda X}\\
    &= 2e^{i\lambda w_k}L_k^\dagger L_k
    - 2 L_k^\dagger L_k,
\end{align*}
where we have used Eqs.~\eqref{eq:conseqB}, \eqref{eq:conseqC}, and \eqref{eq:conseqD} in the last line. 
Therefore, we find 
\begin{align}
    \gen_{\TMH}(\lambda,t) 
    =\frac{i}{2}\tr\Big(\big\{[H,e^{i\lambda X}],e^{-i\lambda X}\}\rho(t)\Big)
    +\gen_{\jump}(\lambda,t). 
\end{align}
Rewriting the first term by the eigenbasis of $X$, we obtain Eq.~\eqref{eq:g_difference}. 

For example, let us imagine a spin system with Hamiltonian $H=\mu S_z$ and jump operators $L_\pm=\sqrt{\gamma_\pm}S_\pm$ ($S_\pm=S_x\pm iS_y$).
If $X$ is the magnetization $S_z$, Eq.~\eqref{eq:commutation} is satisfied as $[S_z,S_\pm]=\pm S_\pm$; since it commutes with the Hamiltonian, $\gen_t(\lambda)$ and $\gen_t^{\fcs}(\lambda)$ coincide, i.e., the stastics is identical for the FCS and the TMH quasiprobability. 
On the other hand, if $X$ is a spin in another direction, such as $S_x$ or $S_y$, or given as $|S_z|=\sqrt{S_z^2}$, Eq.~\eqref{eq:commutation} is no longer fulfilled. Therefore, the change of $X$ is not captured by the FCS. 
On the other hand, the TMH quasiprobability can still describe the fluctuations of $X$ because it can deal with any Hermitian operator.

\clearpage
\appendix
\onecolumngrid
\begin{center}
    \textbf{Supplemental Material}
\end{center}
\bigskip
\twocolumngrid
\setcounter{equation}{0}
\renewcommand{\theequation}{S\arabic{equation}}
\section{Quantization of generating function}
We show that the moment generating function~\eqref{eq:tmh_genfunc} is obtained by appropriately quantizing the classical counterpart. 
Consider classical Markov processes with discrete mesostates $i=1,2,\dots,n$. 
The system's state is described by a probability density $\bm{p}=(p_i)\in \mathbb{R}_{\geq 0}^n$ such that $p_i\geq 0$ and $\sum_ip_i=1$. 
Its time evolution is generally written as the classical master equation 
\begin{align}
    \frac{d\bm{p}(t)}{dt}=\lioucl\big(\bm{p}(t)\big)
    \coloneqq R\bm{p}(t), \label{eq:cme}
\end{align}
where $R$ is the rate matrix, which we assume time-independent~\cite{gardiner2009stochastic}.  
Given the initial probability density $\bm{p}_0$, the classical master equation~\eqref{eq:cme} is solved by $\bm{p}(t)=e^{Rt}\bm{p}_0=e^{\lioucl t}\bm{p}_0$. 

Let us consider a quantity that depends on the system's mesostate, $\bm{f}=(f_i)$. 
Its expectation value at time $t$ is provided as $\langle\bm{f}\rangle_t=\sum_i f_ip_i(t)$. 
By using the standard inner product $\langle\bm{v},\bm{w}\rangle\coloneqq\sum_i v_iw_i$, it is rewritten as $\langle\bm{f}\rangle_t=\langle\bm{f},\bm{p}(t)\rangle$. 
The Heisenberg picture can be discussed by defining $\bm{f}(t)=e^{R^{\trp}t}\bm{f}$, where $\trp$ indicates transpose. 
Because $(e^{Rt})^\trp=e^{R^\trp t}$, we find $\langle \bm{f}(t)\rangle_0=\langle\bm{f}\rangle_t$. 
The adjoint of $\lioucl$ with respect to $\langle\cdot,\cdot\rangle$ is given by $\lioucl^\dagger(\bm{f})=R^\trp\bm{f}$. 
It determines the time evolution of $\bm{f}(t)$ as $\frac{d}{dt}\bm{f}(t)=\lioucl^\dagger(\bm{f}(t))$. 

We next consider the statistics of the increment of $\bm{f}$. 
The joint probability that the system is in state $i$ at time $t$ and in $j$ at time $t+\Delta t$ is given by $[e^{R\Delta t}]_{ji}p_i(t)$~\cite{gardiner2009stochastic}. 
Therefore, the change in $\bm{f}$, denoted by $\Delta\bm{f}$, has the moments 
\begin{align}
    \big\langle(\Delta\bm{f})^n\big\rangle_{t,t+\Delta t}
    = \sum_{i,j}(f_j-f_i)^n[e^{R\Delta t}]_{ji}p_i(t). \label{eq:mom_cl}
\end{align}
As shown later, they are generated by the moment generating function defined as 
\begin{align}
    \Gen_t^\cl(\lambda,\Delta t) 
    \coloneq 
    \Big\langle
    e^{\lioucl^\dagger \Delta t}(e^{i\lambda\bm{f}})
    e^{-i\lambda\bm{f}}
    \Big\rangle_t, \label{eq:generating_cl}
\end{align}
where the exponential and product of vectors are interpreted entrywise; $e^{\bm{v}}=(e^{v_i})$ and $\bm{v}\bm{w}=(v_iw_i)$ for any vectors $\bm{v}$ and $\bm{w}$.
That is, we can prove 
\begin{align}
    \big\langle(\Delta\bm{f})^n\big\rangle_{t,t+\Delta t}
    =(-i\partial_\lambda)^n\Gen_t^\cl(\lambda,\Delta t)|_{\lambda=0}. \label{eq:mom_gen_cl}
\end{align}
Importantly, the TMH moment generating function~\eqref{eq:tmh_genfunc} is led to by the following replacement 
\begin{align}
    \lioucl\to\liou,\;\;
    \bm{f}\to X,\;\; \bm{p}\to \rho, \label{eq:replacement}
\end{align} 
and the introduction of the anticommutator between $e^{\liou^\dagger\Delta t}(e^{i\lambda X})$ and $e^{-i\lambda X}$. 
If the anticommutator is not applied, $\Gen_t^\cl$ will become the moment generating function corresponding to the Kirkwood--Dirac quasiprobability~\cite{gherardini2024quasiprobabilities}. 

Let us prove Eq.~\eqref{eq:mom_gen_cl}. 
First, expand $\Gen_t^\cl$ as 
\begin{align*}
    &\Gen_t^\cl(\lambda,\Delta t) \\
    &=\sum_{m=0}^\infty\sum_{l=0}^\infty 
    \frac{(i\lambda)^m}{m!}\frac{(i\lambda)^l}{l!}
    \big\langle e^{\liou^\dagger \Delta t}(\bm{f}^m)(-\bm{f})^l\big\rangle_t\\
    &=\sum_{r=0}^\infty\sum_{l=0}^r
    \frac{(i\lambda)^r}{r!}\binom{r}{l}
    \big\langle e^{\liou^\dagger \Delta t}(\bm{f}^{r-l})(-\bm{f})^l\big\rangle_t. 
\end{align*}
Differentiating by $\lambda$ for $n$ times and making $\lambda$ zero, we finally find the terms corresponding to $r=n$ in the summation; so we get 
\begin{align*}
    &(-i\partial_\lambda)^n\Gen_t^\cl(\lambda,\Delta t)|_{\lambda=0}\\
    &= \sum_{l=0}^n \binom{n}{l}\big\langle e^{\liou^\dagger \Delta t}(\bm{f}^{n-l})(-\bm{f})^l\big\rangle_t \\
    &= \sum_{l=0}^n \binom{n}{l}
    \sum_{i,j}[e^{R\Delta t}]_{ji}f_j^{n-l}(-f_i)^lp_i(t)\\
    &=\sum_{i,j}[e^{R\Delta t}]_{ji}p_i(t)
    \sum_{l=0}^n \binom{n}{l}f_j^{n-l}(-f_i)^l\\
    &=\sum_{i,j}(f_j-f_i)^n[e^{R\Delta t}]_{ji}p_i(t)
    =\langle(\Delta\bm{f})^n\rangle_{t,t+\Delta t},  
\end{align*}
which concludes the proof. 

Let us focus on the second moment in the short-time limit. 
By expanding $e^{R\Delta t}$, we find 
\begin{align}
    \big\langle(\Delta \bm{f})^2\big\rangle_{t,t+\Delta t}
    =\sum_{i,j}(f_j-f_i)^2R_{ji}p_i(t)\Delta t+O(\Delta t^2), 
\end{align}
where the zeroth order term vanishes because $(f_j-f_i)^n\delta_{ji}p_i=0$ for any $i$ and $j$. 
Thus, the short-time second moment $m_{\bm{f}}^{\cl}(t)=\lim_{\Delta t \to0}\big\langle(\Delta \bm{f})^2\big\rangle_{t,t+\Delta t}/\Delta t$ reads 
\begin{align}
    m_{\bm{f}}^{\cl}(t)
    &=\sum_{i,j}(f_j-f_i)^2R_{ji}p_i(t)\\
    &=\sum_{i,j}f_j^2R_{ji}p_i(t)
    -2\sum_{i,j}f_if_jR_{ji}p_i(t),
\end{align}
where we used $\sum_jR_{ji}=0$. 
By remembering $\liou_\cl^\dagger =R^{\trp}$, we can further rewrite this equation into 
\begin{align}
    m_{\bm{f}}^{\cl}(t)
    =\big\langle\liou_\cl^\dagger (\bm{f}^2)-2\liou_\cl^\dagger (\bm{f})\bm{f}\big\rangle_t.  
\end{align}
We can quantize this equation and recover Eq.~\eqref{eq:mxexpression} by $\liou^\dagger (\bm{f})\bm{f}\to\{\liou^\dagger (X),X\}/2$ in addition to the replacement in Eq.~\eqref{eq:replacement}. 

\section{Derivation of Eq.~\eqref{eq:tur_old}}
We concisely review the derivation of Eq.~\eqref{eq:tur_old} in Ref.~\cite{yoshimura2025force}. 
For the proof of unproven facts, please see that paper. 

The fundamental strategy is as follows: 1) rewrite the entropy production rate as an inner product, 2) rewrite it as a squared norm, and 3) apply the Cauchy--Schwarz inequality. 
First, notice that for each pair of jumps $(k,-k)$, we can take coefficients $\gamma_{\pm k}>0$ and operators $\tilde{L}_{\pm k}$ such that $L_{\pm k}=\sqrt{\gamma_{\pm k}}\tilde{L}_{\pm k}$ and $\tilde{L}_{-k}=\tilde{L}_k^\dagger$. The local detailed balance imposes $\gamma_k/\gamma_{-k}=e^{s_k}$ ($k_\mathrm{B}$ is set to be unity). 
Then, we define 
\begin{align}
    \curr_k(\rho)&\coloneqq \begin{pmatrix}
        O&J_{-k}(\rho)\\ J_k(\rho)& O
    \end{pmatrix},\\
    \force_k(\rho)&\coloneqq \begin{pmatrix}
        O&F_{-k}(\rho)\\ F_k(\rho)& O
    \end{pmatrix},
\end{align}
where 
\begin{align}
    J_k(\rho)&\coloneqq \frac{1}{2}(\gamma_k \tilde{L}_k\rho- \gamma_{-k}\rho \tilde{L}_{k}),\\
    F_k(\rho) &\coloneqq s_k\tilde{L}_k + [\tilde{L}_k,\ln\rho]. 
\end{align}
Now, $\curr_k(\rho)$ and $\force_k(\rho)$ are anti-Hermitian operators on $\mathbb{C}^2\otimes\hilb$, where $\hilb$ is the original Hilbert space. 
By aligning them, we further define 
\begin{align}
    \curr(\rho)\coloneqq \bigoplus_k\curr_k(\rho),\quad
    \force(\rho)\coloneqq \bigoplus_k\force_k(\rho), 
\end{align}
where the direct sum is taken over the pairs of jumps (thus, if $k$ is counted, $-k$ is not). 
They become operators on $\hilbb\otimes\hilb$, where $\hilbb$ is a complex vector space whose dimension is equal to the number of jumps. 
Then, the entropy production rate (EPR) defined in Eq.~\eqref{eq:epr} is given by 
\begin{align}
    \epr(\rho) = \langle \curr(\rho), \force(\rho) \rangle, 
\end{align}
where $\langle\cdot,\cdot\rangle$ is the Hilbert--Schmidt inner product, defined as $\langle A, B\rangle=\tr(A^\dagger B)$. 

Next, we introduce a super-operator $\nabla_{\strc}$ that maps operators on $\hilb$ to those on $\hilbb\otimes\hilb$. 
It is defined by 
\begin{align}
    \nabla_{\strc}A \coloneqq [I_{\hilbb}\otimes A,\strc], 
\end{align}
where $I_{\hilbb}$ is the identity operator on $\hilbb$ and 
\begin{align}
    \strc\coloneqq \bigoplus_k
    \begin{pmatrix}
        O & \tilde{L}_{-k}\\ \tilde{L}_k & O
    \end{pmatrix}. 
\end{align}
It acts as a gradient, and its adjoint turns $\curr(\rho)$ into the dissipator as $\nabla_{\strc}^\dagger \curr(\rho)=\diss(\rho)$. 

Finally, for a positive operator $G$, we introduce a super-operator 
\begin{align}
    \ons_{G}(A)\coloneqq \int_0^1 G^{s}XG^{1-s}ds, 
\end{align}
which satisfies $\ons_{\Gamma\otimes\rho}(\force(\rho))=\curr(\rho)$ with 
\begin{align}
    \Gamma=\bigoplus_k \begin{pmatrix}
        \gamma_k/2&0\\0&\gamma_{-k}/2
    \end{pmatrix}. 
\end{align}
We can define an inner product $\langle A,B\rangle_G = \langle A,\ons_G(B)\rangle$ and the induced norm $\lVert A\rVert_G=\sqrt{\langle A,A\rangle_G}$, which allows us to rewrite the EPR as  
\begin{align}
    \epr(\rho)
    =\lVert \force(\rho)\rVert_{\Gamma\otimes\rho}^2. 
\end{align}
By applying the Cauchy--Schwarz inequality, we get 
\begin{align}
    \epr(\rho)\lVert \nabla_{\strc}X\rVert_{\Gamma\otimes\rho}^2
    \geq |\langle \force(\rho),\nabla_{\strc}X\rangle_{\Gamma\otimes\rho}|^2. 
\end{align}
Because $\ons_{\Gamma\otimes\rho}(\force(\rho))=\curr(\rho)$ holds, the right-hand side leads to $\tr(X\diss(\rho))$. 
On the other hand, we can generally show $\langle A,\ons_G(A)\rangle \leq \frac{1}{2}\tr(A^\dagger \{G,A\})$; by a direct calculation, we can derive that this upper bound leads to $\diff_X(\rho)$ when $A=\nabla_{\strc}X$ and $G=\Gamma\otimes\rho$ and obtain Eq.~\eqref{eq:tur_old}. 

\section{Computational details of Example}
We provide the details of the computation in the example in the main text. 
We aim to compute $\mathcal{T}_{s's}(\rho)$ and $\bar{\lambda}(\rho)$ for $\rho=\rho^\pm$ and show $m_H(\rho^+)=O(N^2)$ and $m_H(\rho^-)=O(1)$. 
Further, we also compute the current $J_H^{\mathrm{d}}(\rho^-)$ and the EPR $\epr(\rho^-)$ and confirm the inequality.
Here, the model we consider consists of Hamiltonian $H=\omega\sum_j\ketbras{e,j}$ and jump operators $L_+=\sqrt{\gamma_+}\sum_{j,j'}\ketbrad{e,j}{g,j'}$ and $L_-=\sqrt{\gamma_-}\sum_{j,j'}\ketbrad{g,j}{e,j'}$. 
The states of interest $\rho^\pm$ are generated as $\rho^\pm = \sum_{s=g,e}p_s\ketbras{s,\pm}$, where $\ket{s,\pm}=\frac{1}{\sqrt{N}}\sum_j (\pm 1)^j\ket{s,j}$ and $p_s$ satisfies $p_s\geq 0$ and $p_g+p_e=1$. 

Because now we consider $X=H$, we need to compute 
\begin{align}
    T_{s'j'sj}(\rho)=
    \frac{1}{2}\tr\big(\{\diss^\dagger \Pi_{s'j'},\Pi_{sj}\}\rho\big). 
\end{align}
For convenience, we define 
\begin{align}
    \sigma_{s,j,j'}&\coloneqq 
    \frac{1}{2}\big(\ketbrad{s,j}{s,j'}+\ketbrad{s,j'}{s,j}\big),\\
    \Lambda_{s,j}&\coloneqq 
    \sum_{j'}\sigma_{s,j,j'}. 
\end{align}
They satisfy 
\begin{gather}
    \sigma_{s,j,j'}=\sigma_{s,j',j},\;\;
    \sigma_{s,j,j}=\Pi_{s,j},\\
    \tr(\sigma_{s,j,j'}\rho)=\mathrm{Re}\bra{s,j}\rho\ket{s,j'},\\
    \sum_{j}\Lambda_{s,j}
    = \sum_{j,j'}\ketbrad{s,j}{s,j'},\\
    \frac{1}{2}\sum_{j'}\{\Lambda_{s',j'},\Pi_{s,j}\}
    =\delta_{s,s'}\Lambda_{s,j}. 
\end{gather}
By using them, we obtain 
\begin{align}
    L_+^\dagger \Pi_{s,j} L_+
    &= \gamma_+ \delta_{s,e}\sum_{j'}\Lambda_{g,j'},\\
    L_-^\dagger \Pi_{s,j} L_-
    &= \gamma_- \delta_{s,g}\sum_{j'}\Lambda_{e,j'}, 
\end{align}
and 
\begin{align}
    L_+^\dagger L_+
    &= \gamma_+N\sum_{j'}\Lambda_{g,j'},\\
    L_-^\dagger L_-
    &= \gamma_-N\sum_{j'}\Lambda_{e,j'}. 
\end{align}
Then, we find 
\begin{align}
    \diss^\dagger \Pi_{s,j}
    &= \delta_{s,e}\bigg(\gamma_+\sum_{j'}\Lambda_{g,j'}
    -\gamma_-N\Lambda_{e,j}\bigg)\notag\\
    &+\delta_{s,g}\bigg(\gamma_-\sum_{j'}\Lambda_{e,j'}
    -\gamma_+N\Lambda_{g,j}\bigg). 
\end{align}
Consequently, we get 
\begin{align}
    &\frac{1}{2}\{\diss^\dagger \Pi_{s',j'},\Pi_{s,j}\}\notag\\
    &= \delta_{s',e}\delta_{s,g}\gamma_+\Lambda_{g,j}
    - \frac{\delta_{s',e}\delta_{s,e}}{2}\gamma_-N(\delta_{j,j'}\Lambda_{e,j}
    +\sigma_{e,j,j'})\notag\\
    &+ \delta_{s',g}\delta_{s,e}\gamma_-\Lambda_{e,j}
    - \frac{\delta_{s',g}\delta_{s,g}}{2}\gamma_+N(\delta_{j,j'}\Lambda_{g,j}
    +\sigma_{g,j,j'}). 
\end{align}
For density matrix $\rho$, the fluxes are given as 
\begin{align}
    T_{ej'gj}(\rho)
    &= \gamma_{+}\sum_{j''}\mathrm{Re}\bra{g,j}\rho\ket{g,j''}\\
    T_{gj'ej}(\rho)
    &= \gamma_{-}\sum_{j''}\mathrm{Re}\bra{e,j}\rho\ket{e,j''}\\
    T_{gj'gj}(\rho)
    &= -\frac{\gamma_+N}{2}
    \bigg(\mathrm{Re}\bra{g,j}\rho\ket{g,j'}\notag\\
    &\phantom{===}+\delta_{j,j'}\sum_{j''}\mathrm{Re}\bra{g,j}\rho\ket{g,j''}\bigg)\\
    T_{ej'ej}(\rho)
    &= -\frac{\gamma_-N}{2}
    \bigg(\mathrm{Re}\bra{e,j}\rho\ket{e,j'}\notag\\
    &\phantom{===}+\delta_{j,j'}\sum_{j''}\mathrm{Re}\bra{e,j}\rho\ket{e,j''}\bigg). 
\end{align}
For $\rho=\rho^+$ and $\rho^-$, because $\bra{s,j}\rho^+\ket{s,j''}=p_s/N$ and $\bra{s,j}\rho^-\ket{s,j''}=(-1)^{j+j''}p_s/N$, we find 
\begin{align}
    T_{ej'gj}(\rho^+)
    = p_g\gamma_+,\;\;
    T_{ej'gj}(\rho^-)
    = (-1)^{j+1}p_g\gamma_+\frac{\chi(N)}{N}, 
\end{align}
where we used $\sum_{j}(-1)^j=-\chi(N)$. 
Thus, 
\begin{align}
    \mathcal{T}_{eg}(\rho^+)
    = p_g\gamma_+N^2,\quad 
    \mathcal{T}_{eg}(\rho^-)
    = p_g\gamma_+ \chi(N). \label{eq:tegplus}
\end{align}
Note that $\chi(N)^2=\chi(N)$. 
We also see 
\begin{align}
    T_{gj'gj}(\rho^+)
    &= -\frac{\gamma_+p_g}{2}(1+\delta_{j,j'}N)\\
    T_{gj'gj}(\rho^-)
    &= -\frac{(-1)^j\gamma_+p_g}{2}
    ((-1)^{j'}-\delta_{j,j'}\chi(N)). 
\end{align}
Therefore, we can get  
\begin{align}
    \mathcal{T}_{gg}(\rho^+)
    &= -\frac{\gamma_+p_g}{2}N(N-1),\\
    \mathcal{T}_{gg}(\rho^-)
    &= \frac{\gamma_+p_g}{2}(N-\chi(N)) 
\end{align}
because 
\begin{align}
    \sum_{j,j'(\neq j)}(-1)^j(-1)^{j'}
    &= \bigg(\sum_j(-1)^j\bigg)^2 - \sum_{j}(-1)^{2j}\\
    &= \chi(N) - N.
\end{align}
The other integrated fluxes are given in the same way. 
Besides, we can compute the average escape rates as 
\begin{align}
    \bar{\lambda}(\rho^+)
    &= \frac{\gamma_+p_g+\gamma_-p_e}{2}N(N+1),\\
    \bar{\lambda}(\rho^-)
    &= \frac{\gamma_+p_g+\gamma_-p_e}{2}(N+\chi(N)). 
\end{align}
Finally, from Eq.~\eqref{eq:tegplus}, we find 
\begin{align}
    m_H(\rho^+)
    &= \omega^2(\gamma_+p_g+\gamma_-p_e)N^2=O(N^2),\\
    m_H(\rho^-)
    &= \omega^2(\gamma_+p_g+\gamma_-p_e)\chi(N)=O(1).
\end{align}

The current $J_H^{\mathrm{d}}(\rho^-)$ is provided as
\begin{align}
    J_H^{\mathrm{d}}(\rho^-)
    &=\omega(\tr(L_+^\dagger L_+\rho^-)-\tr(L_-^\dagger L_-\rho^-))\\
    &=\omega(\gamma_+p_g-\gamma_-p_e)\chi(N). 
\end{align}
On the other hand, the EPR becomes zero if $\chi(N)=0$, or diverges otherwise at $\rho^-$; when $\chi(N)=0$, we have $\diss(\rho^-)=0$ because 
\begin{align}
    \braketd{e,+}{e,-}= \braketd{g,+}{g,-} = \frac{1}{N}\chi(N)
\end{align}
and the jump operators can read $L_+=N\sqrt{\gamma_+}\ketbrad{e,+}{g,+}$ and $L_-=N\sqrt{\gamma_-}\ketbrad{g,+}{e,+}$, regardless of the value of $\chi(N)$.
In a detailed balanced system, such as the one under consideration, the EPR can be rewritten as~\cite{breuer2002theory} 
\begin{align}
    \epr(\rho)=\tr(\diss(\rho)(-\ln\rho+\beta H)). 
\end{align}
Therefore, the EPR vanishes when $\rho=\rho^-$ and $\chi(N)=0$. 
This does not contradict the TUR~\eqref{eq:qtur} because if $\diss(\rho)=0$, the dissipative current $J_X^{\mathrm{d}}(\rho)$ vanishes for any $X$.
Furthermore, the equality $m_H(\rho^-)=0$, which occurs when $\chi(N)=0$, is also consistent with the fact that both $\epr(\rho^-)$ and $J_H^{\mathrm{d}}(\rho)$ are zero.

On the other hand, the EPR diverges when $\chi(N)$ is nonzero (i.e., when $N$ is odd). 
This is proved from the following general expression with the eigendecomposition $\rho=\sum_ip_i\ketbras{i}$~\cite{funo2019speed}:
\begin{align}
    \epr(\rho)=\sum_{i<j}\sum_k (R_{ji}^kp_i-R_{ij}^{-k}p_i)\ln\frac{R_{ji}^kp_i}{R_{ij}^{-k}p_j}. \label{eq:app-81}
\end{align}
Here, note that each term in the sum is individually non-negative. 
Now, $\rho^-$ has a nontrivial null space, which can contain $\ket{e_{12}}=(\ket{e,1}+\ket{e,2})/\sqrt{2}$ (note $\braketd{e_{12}}{e,-}=\braketd{e_{12}}{g,-}=0$). 
Then, $R_{e_{12},(g,-)}^+$ and $R_{(g,-),e_{12}}^-$ can be nonzero because 
\begin{align}
    R_{e_{12},(g,-)}^+&=|\bra{e_{12}}L_+\ket{g,-}|^2
    = \frac{2\gamma_+}{N^3}\chi(N),\\
    R_{(g,-),e_{12}}^-&=|\bra{g,-}L_-\ket{e_{12}}|^2
    = \frac{2\gamma_-}{N^3}\chi(N). 
\end{align}
Therefore, if $\chi(N)\neq 0$, 
\begin{align}
    (R_{e_{12},(g,-)}^+p_g-R_{(g,-),e_{12}}^-p_{e_{12}})\ln\frac{R_{e_{12},(g,-)}^+p_g}{R_{(g,-),e_{12}}^{-}p_{e_{12}}}
    =\infty
\end{align}
because $p_{e_{12}}=\bra{e_{12}}\rho\ket{e_{12}}=0$. 
Since each term in Eq.~\eqref{eq:app-81} is non-negative, the divergence of a single term means that of the sum. 
The divergence occurs because now $\diss(\rho^-)$ is not zero and some state transitions in the dissipative dynamics (such as the one between $(g,-)$ and $e_{12}$) become absolutely irreversible. 
Since the other quantities, namely, the dissipative current and the short-time fluctuation, are insensitive to such irreversibility, they remain finite even though the EPR diverges. This is again consistent with the TUR~\eqref{eq:qtur}.

\section{Hierarchy of conditions}
We compare the two conditions of non-classicality, the one based on coherence and the other given by quasiprobabilities. 
Assume that $\{\ket{s,j}\}$ is a classical eigenbasis; i.e., $|\bra{s',j'}L_k\ket{s,j}|$ and the number of jump $k$ such that $|\bra{s',j'}L_k\ket{s,j}|\neq 0$ remain $O(1)$ as $N$ increases. 
For simplicity, we use Greek letters such as $\alpha$ and $\beta$ to indicate pair labels, e.g., $(s,j)$. 
The matrix elements with respect to the basis are denoted by $A_{\alpha\beta}$ (i.e., $A_{\alpha\beta}=\bra{\alpha}A\ket{\beta}$).

For the given basis, 
we can split $\rho$ as $\rho=\rho_d+\chi$ with the diagonal part $\rho_d=\sum_{\alpha}\Pi_\alpha\rho\Pi_\alpha$ and the off-diagonal part $\chi=\rho-\rho_d$. 
We say $\rho$ satisfies the coherence condition (Co) if $\lVert\chi\rVert_{l^1}\to \infty$ holds when  $N\rightarrow\infty$, where $\lVert\chi\rVert_{l^1}\coloneqq\sum_{\alpha,\beta}|\chi_{\alpha\beta}|$.
This condition can be rephrased as $C_{l^1}(\rho)\to\infty$, where $C_{l^1}(\rho)\coloneqq\sum_{\alpha,\beta(\neq\alpha)}|\rho_{\alpha\beta}|$ is the so-called $l^1$-norm of coherence. Because $\chi$ has no nonzero diagonal elements, we have $\lVert\chi\rVert_{l^1}=C_{l^1}(\chi)$, which is also equal to $C_{l^1}(\rho)$.

We prove that the violation of (Co) (i.e., $\lVert\chi\rVert_{l^1}=O(1)$) implies the violation of both (Q1) and (Q2), which means that the quasiprobability-based conditions are stricter than the coherence condition as formulated above. 
First, we calculate $T_{\alpha'\alpha}(\rho)=\frac{1}{2}\tr\big(\{\liou^\dagger\Pi_{\alpha'},\Pi_\alpha\}\rho\big)$. 
Let us define
\begin{align}
    S_{\alpha\beta\gamma}\coloneqq \sum_k\bra{\alpha}L_k^\dagger\ketbras{\beta}L_k\ket{\gamma}. 
\end{align}
It is worth mentioning that $S_{\alpha\beta\alpha}=R_{\beta\alpha}$. 
Then, we obtain 
\begin{align}
    &\tr\big((\liou^\dagger\Pi_{\alpha'})\Pi_\alpha\rho\big)
    = \sum_\beta S_{\beta\alpha'\alpha}\rho_{\alpha\beta}\notag\\
    &\quad\quad-\frac{1}{2}\sum_\gamma S_{\alpha'\gamma\alpha}\rho_{\alpha\alpha'}
    -\frac{1}{2}\delta_{\alpha\alpha'}\sum_{\beta,\gamma} S_{\beta\gamma\alpha}\rho_{\alpha\beta}. \label{eq:app-78}
\end{align}
Because $\tr\big(\Pi_\alpha(\liou^\dagger\Pi_{\alpha'})\rho\big)=\tr\big((\liou^\dagger\Pi_{\alpha'})\Pi_\alpha\rho\big)^*$, $T_{\alpha'\alpha}(\rho)$ will be presented as the real part of Eq.~\eqref{eq:app-78}.

Next, we discuss the effect from $\chi$ to $\mathcal{T}_{s's}(\rho)$ and $\bar{\lambda}(\rho)$. 
Due to the linearity of $T_{\alpha'\alpha}(\rho)$, $\mathcal{T}_{s's}(\rho)$ and $\bar{\lambda}(\rho)$ are decomposed as 
\begin{align}
    \mathcal{T}_{s's}(\rho)&=\mathcal{T}_{s's}(\rho_d)+\mathcal{T}_{s's}(\chi),\\
    \bar{\lambda}(\rho)&=\bar{\lambda}(\rho_d)+\bar{\lambda}(\chi). 
\end{align}
As explained in the main text, $T_{\alpha'\alpha}(\rho_d)\geq 0$ when $\alpha\neq \alpha'$, and $\bar{\lambda}(\rho_d)=O(N)$ because $\rho_d$ and $X$ commute and are simultaneously diagonalized by a classical eigenbasis. 
Thus, to prove the statement, it suffices to show $|\mathcal{T}_{s's}(\chi)|=O(N)$ and $\bar{\lambda}(\chi)=O(N)$. 
These are proven as follows: first, define $[s]$ as the set $\{(s,j)\}_{j=1}^N$. Then, we have
\onecolumngrid
\begin{align}
    |\mathcal{T}_{s's}(\chi)|
    &\leq \sum_{\alpha'\in[s']}\sum_{\alpha(\neq\alpha')\in[s]}
    |T_{\alpha'\alpha}(\chi)|
    \leq \sum_{\alpha'\in[s']}\sum_{\alpha(\neq\alpha')\in[s]}
    \bigg(\sum_\beta |S_{\beta\alpha'\alpha}||\chi_{\alpha\beta}|
    +\frac{1}{2}\sum_{\gamma}|S_{\alpha'\gamma\alpha}||\chi_{\alpha\alpha'}|\bigg)\\
    &\leq \sum_{\alpha'\in[s']}\max_{\alpha,\beta}|S_{\beta\alpha'\alpha}|\lVert\chi\rVert_{l^1}
    +\frac{1}{2}\sum_\gamma\max_{\alpha,\alpha'}|S_{\alpha'\gamma\alpha}|\lVert\chi\rVert_{l^1}, 
\end{align}
where we used the triangle inequality and the H\"{o}lder inequality $\sum_i|x_iy_i|\leq \max_i|x_i|\sum_i|y_i|$. 
The classicality condition tells that $\max_{\alpha,\beta}|S_{\beta\alpha'\alpha}|=O(1)$ for any $\alpha'$. 
Since the dummy variables in the summations ($\alpha'$ and $\gamma$) run sets that have $O(N)$ components, $|\mathcal{T}_{s's}(\chi)|$ is bounded by a quantity of order $N$. 
Similarly, $|\bar{\lambda}(\chi)|$ is bounded by an $O(N)$ quantity because 
\begin{align}
    |\bar{\lambda}(\chi)|
    &\leq \sum_{\alpha}|T_{\alpha\alpha}(\chi)|
    \leq \sum_{\alpha,\beta} |S_{\beta\alpha\alpha}||\chi_{\alpha\beta}|
    + \frac{1}{2}\sum_{\alpha,\beta,\gamma}|S_{\beta\gamma\alpha}||\chi_{\alpha\beta}|\\
    &\leq \max_{\alpha,\beta}|S_{\beta\alpha\alpha}|\lVert\chi\rVert_{l^1}
    +\frac{1}{2}\sum_\gamma\max_{\alpha,\beta}|S_{\beta\gamma\alpha}|\lVert\chi\rVert_{l^1}. 
\end{align}

\end{document}

%% file: biblio.bbl
%